\documentstyle[12pt,epsf]{article}
\begin{document}
\begin{titlepage}
\begin{center}
\vspace{15mm}
\Large
%
Dynamics of a planar domain wall with oscillating thickness in $\lambda \Phi^{4}$ model \\
\vspace{15mm} by \\
\vspace{8mm} 
\large 
Maciej \'Slusarczyk${}^{\dag}$ \\
\vspace{15mm}  
\normalsize
Jagiellonian University, Institute of Physics \\
Reymonta 4, 30-059 Cracow, Poland \\
\vspace{20mm}
%
%
\large
\begin{abstract}
Domain wall - type solution with oscillating thickness in a real, scalar field model 
is investigated with the help of a polynomial approximation. We propose a simple extension of the
polynomial approximation method. In this approach we calculate higher order corrections to the 
planar domain wall solution, find that the domain wall with oscillating thickness radiates,
and compute dumping of oscillations of the domain wall. 
\end{abstract}
\vspace{10mm}  
\end{center}
\normalsize
\vspace{1.4cm}
\noindent
\underline{\hspace*{16cm}}
${}^{\dag}$ e-mail: mslus@thrisc.if.uj.edu.pl
\end{titlepage}
%
%
\section{Introduction}
Topological defects constitues an important class of solutions in field-theoretical models with 
degenerate vacua. They play very important role in several branches of physics. Let us mention here
field-theoretical cosmology and the cosmic strings hypothesis (see \cite{Kibble}, \cite{Vilenkin}, \cite{Zurek}),
dynamics of superconductors, superfluids and liquid crystals in condensed matter physics (see
\cite{Zurek} - \cite{l_cr1}) as well as a flux tube in QCD. \\
This short and incomplete list shows the necessity of having effective computing methods to study the dynamics
of topological defects. In spite of the increasing development of mathematical techniques to
solve nonlinear equations, exact solutions seem not to be the rule and numerical methods have
been the most common approach to study properties of topological defects. Therefore analytical and
perturbative methods are of great interest and importance.

In this paper we study domain wall type solution in the $ \lambda \Phi^{4} $ model. In paper
\cite{Pelka} excited domain wall of this kind was considered and the radiation emitted from the
domain wall was found. In our approach we further develop the method of a so-called polynomial
approximation (see \cite{dwall}, \cite{dwall1}), which is used to construct the domain wall type
solution with time-dependent thickness. We show how to compute corrections to the 
polynomial solution. The corrections consist of two parts: the static one and the time-dependent one. The 
time-dependent part of the correction contains radiation emitted by domain wall with oscillating
thickness.

The plan of our paper is the following. In the next section we present the model. In section
\ref{pa} we derive the time-dependent planar domain wall
solution in the polynomial approximation. Section \ref{corr} is devoted to the detailed analysis of this 
solution. We present the method of finding correction to the polynomial solution. In sections
\ref{static} and \ref{dyn} we calculate static and time-dependent part of this correction.
In section \ref{back} we analyze the backreaction of the radiation emitted on the domain wall. In
section \ref{sum} we shortly summarize the main points of our work.
%
%
\section{The Model}
We consider the model with single, scalar, real - valued field $ \Phi $, defined by the action:
\begin{equation}
{\cal S} = \int d^{4}x [-\frac{1}{2} \eta_{ \mu \nu} \partial^{\mu} \Phi \partial^{\nu}
\Phi - \frac{ \lambda }{2} (\Phi^{2}-v^{2})^{2}],                 \label{lagr}
\end{equation}
where $\eta_{\mu \nu}=diag(-1,1,1,1)$ and $\lambda$, $v$ are positive constants. The corresponding 
equation of motion for the field $\Phi$ has the form:
\begin{equation}
\partial_{\mu} \partial^{\mu} \Phi - 2 \lambda \Phi ( \Phi^{2}-v^{2}) = 0.  \label{rr}
\end{equation}
The energy functional for the model is given by the formula:
\begin{equation}
E[ \Phi ] = \frac{1}{2} \int d^{3} \vec{x}[ \partial_{\mu} \Phi \partial^{\mu} \Phi +
\lambda( \Phi^{2}-v^{2})^{2}].                                
\label{energia}
\end{equation}
It is convenient to rescale the space - time coordinates and the scalar field as follows:
\begin{eqnarray}\phi & = & \frac{\Phi}{v},  \nonumber \\
t & = & \alpha x^{0},       \nonumber \\
\xi & = & \alpha x^{3},     \nonumber \\
\lefteqn{ \tilde{x}^{1} = \alpha x^{1}, \; \;  \tilde{x}^{2} = \alpha x^{2},} 
\end{eqnarray}
where $ \alpha = \sqrt{\lambda v^{2}}$. The new variables are dimensionless.
The vacuum values of $\phi$ in the considered model are equal to $\pm 1$. 
Configuration of the field which smoothly interpolates between these two vacua
is called the domain wall. Our goal is to construct the domain wall configuration, 
localised on the $\tilde{x}^{1}-\tilde{x}^{2}$ plane, with the time-dependent width. 
The planar domain wall distinguishes the direction perpendicular to the wall plane,
given in our case by the coordinate lines of $\xi$. As we consider the configuration
with $\phi$ independent of the coordinates $\tilde{x}^{1}$ and $\tilde{x}^{2}$, we can 
restrict our approach to the $1+1$ dimensional model. Equation (\ref{rr}) then takes the form:     
\begin{equation}
 - \frac{\partial^{2} \phi}{\partial t^{2}} + \frac{\partial^{2} \phi} {\partial \xi^{2}} - 2 \phi (\phi^{2} - 1) = 0.
    \label{prr}
\end{equation}
Static solutions of this equation are well-known. They have the form:
\begin{equation}
\phi(\xi) = \pm \tanh \xi,   \label{stat}
\end{equation}
%
%
\section{Domain wall solution in the polynomial approximation.}
\label{pa}
In this section we construct approximate domain wall type solution of Eq. (\ref{prr})
with time-dependent width. In order to realize this we use the method of a polynomial approximation, 
whose detailed description is given in \cite{dwall}. The basic idea of this approach is to approximate
 the scalar field inside the domain wall by the polynomial in the variable $\xi$ with time-dependent 
coefficients. Thus, inside an interval $[ - \xi_{1}, \xi_{0} ]$ ($\xi_{0}$ and $\xi_{1}$ are 
positive) we have:
\begin{equation}
\phi(t,\xi) = a(t) \xi + \frac{1}{2!} b(t) \xi^{2} + \frac{1}{3!} c(t)\xi^{3}.
\label{ansatz}
\end{equation}
The domain wall solution is characterized by the fact, that for sufficiently large
$\mid \xi \mid$ the field approaches its vacuum values $\pm 1$:
\begin{eqnarray}
\phi(t,\xi) & = &+1 \;\;\; \mbox{    for    } \;\;\;\xi \geq \xi_{0}, \nonumber \\
\phi(t,\xi) & =  & -1  \;\;\; \mbox{    for    } \;\;\; \xi \leq - \xi_{1}. 
\label{proz}
\end{eqnarray}
It is possible (see \cite{dwall}) to adopt more accurate asymptotics for large $\mid \xi \mid$,
with exponential correction $\exp ( -2 \alpha \xi ) $ to the vacuum values $ \pm 1$. For simplicity
 of further calculations we use expressions (\ref{proz}). \\
We can tune accuracy of our approximation changing degree of the polynomial (\ref{ansatz}).
One can easily notice, that the cubic Ansatz presented above is the simplest, nontrivial choice.
The Ansatz (\ref{ansatz}) should be smoothly matched with the vacuum solutions at $\xi = \xi_{0}$ and $\xi = - \xi_{1}$.
The matching conditions follow in a standard manner from Eq. (\ref{prr}). One integrates 
Eq. (\ref{prr}) over $\xi$ in arbitrary small intervals $[- \xi_{1} - \epsilon , - \xi_{1} + \epsilon]$  ; 
$[ \xi_{0} - \epsilon , \xi_{0} + \epsilon]$ and lets $\epsilon \rightarrow 0$. This implies:
\begin{eqnarray}
\partial_{\xi} \phi(t , \xi) \mid_{\xi = \xi_{0}} & = & 0, \nonumber \\
\partial_{\xi} \phi(t , \xi) \mid_{\xi = -\xi_{1}} & = & 0.  \label{zsz1}
\end{eqnarray}
From Eq. (\ref{proz}) one gets:
\begin{eqnarray}
\phi(t,\xi_{0}) & = & +1, \nonumber \\
\phi(t,- \xi_{1}) & = & -1.    \label{zsz2}
\end{eqnarray}
The Ansatz (\ref{ansatz}) with matching conditions (\ref{zsz1}) - (\ref{zsz2}) allows us to find
solution of Eq. (\ref{prr}) in the proposed form. Inserting the expansion (\ref{ansatz}) into 
the matching conditions we get, analogously to the case of cylindrical domain wall discussed 
in \cite{dwall}, the following conditions for functions $a(t)$, $b(t)$, $c(t)$ and parameters 
$\xi_{0}$, $\xi_{1}$: 
\begin{equation}
\xi_{0} = \xi_{1}, \; \; a(t)=\frac{3}{2 \xi_{0}}, \; \; b(t) = 0, \; \; 
c(t) = - \frac{8a^{3}}{9}.     
\label{wr}
\end{equation}
Then we insert the expansion (\ref{ansatz}) into (\ref{prr}) and equate to zero coefficients 
in front of successive powers of $\xi$. After some easy algebra, considering conditions
(\ref{wr}) we get the following equation:
\begin{equation}
\ddot{a} + \frac{8}{9} a^{3} - 2 a = 0.    
\label{rna}
\end{equation}
Function $c(t)$ is related to $a(t)$ (see formula (\ref{wr})) and the last step in our approach is to 
solve Eq. (\ref{rna}). We construct an approximate solution of Eq. (\ref{rna}), which is convenient
 for the further analysis. With variables redefined as follows:
\begin{equation}
a(t) = \frac{3}{2}  A(t), \; \; \;
\tau = 2 t,
\end{equation}
Eq. (\ref{rna}) takes the form:
\begin{equation}
\ddot{A} = A - A^{3},  
\label{rna1}
\end{equation}
where now dot denotes the derivative with respect to $\tau$.
Numerical analysis of Eq. (\ref{rna1}) shows that there exist periodic solutions of two kinds.
 Periodic solutions of the first kind oscillate around static solution $A=1$ and take only 
positive values; periodic solutions of the second kind oscillate around static solution $A=-1$
and take only negative values. We can easily analyse the oscillating solutions of 
Eq. (\ref{rna1}) rewriting it in the form of the two equations of the first order by substitution
$\dot{A} = B$. Then periodic solutions are equivalent to closed trajectories in (A,B) configuration space.
$(1,0)$, $(-1,0)$ are the central critical points.  The (A,B) configuration space is presented in Fig. 1.
\\
In the further part of our discussion we consider only the periodic solutions oscillating around 
$A=1$, because of the connection between $A$ and the domain wall thickness parameter, $\xi_{0}$. 
When $A(t) > 0$ for all $\tau$ we have also $\xi_{0},\xi{1} > 0$, which is consistent with the  
interpretation of these parameters. The periodic solutions oscillating around $A=-1$ 
are symmetric (in the sense that when one changes $\xi_{0}$ and $\xi_{1}$ by each other it has 
no influence on the dynamics of our system), so we don't discuss this situation. Considering all these remarks we restrict
 our investigation to solutions oscillating with small amplitude around $A=1$. We can write:
\begin{equation}
A(\tau) = 1 + \Psi (\tau),   
\label{omega}
\end{equation}
where $\Psi (\tau)$ is small, periodic function. Inserting (\ref{omega}) into Eq. (\ref{rna1}) 
one gets:
\begin{equation}
\ddot{\Psi} + \Psi = - \frac{1}{2} \Psi^{3}- \frac{3}{2} \Psi^{2}.     
\label{rna2}
\end{equation}
Expression (\ref{rna2}) has the form of oscillator equation with nonlinear terms. Our goal is to
find its perturbative solution with initial condtions: $\Psi (0) = \Psi_{0}$, $\dot{\Psi} (0) = 0$, 
which describes initially static, sqeezed wall. One can solve the nonlinear oscillator equation of 
type (\ref{rna2}) using Krylov - Bogolubov method (see \cite{Korn}).The general solution has the
 form:
\begin{equation}
\Psi(\tau) = \omega \cos \varphi(\tau) + \frac{1}{2} \alpha_{0} (\omega) +
\sum_{k \geq 1} \alpha_{k} (\omega) \cos k \varphi (\tau) + \beta_{k} (\omega) \sin k \varphi
 (\tau),                
\label{kr_bog}
\end{equation}
where functions $\omega, \; \varphi, \; \alpha, \; \beta$ can be calculated in a standard manner
 (see \cite{Korn}) with initial condition $\omega(0) = \delta$. In our case we find:
\begin{equation}
\Psi (\tau) = \delta \cos \left[ \left( 1 + \frac{3 \delta^{2}}{16}\right) \tau\right] -
\frac{3 \delta^{2}}{4} + \frac{\delta^{2}}{4} \cos \left[ 2 \left( 1+\frac{2 \delta^{2}}{16} 
\right) \tau \right] + O(\delta^{3}),
\end{equation} 
where $\delta$ is a small parameter. Thus we get solution $a(t)$:
\begin{equation}
a(t) = \frac{3}{2} \left( 1 + \delta \cos \left[ 2 \left( 1 + \frac{3 \delta^{2}}{16} \right)
 t \right] - \frac{3 \delta^{2}}{4} + \frac{\delta^{2}}{4} \cos \left[ 4 \left( 1 + 
\frac{3 \delta^{2}}{16} \right) t \right] \right) + O( \delta^{3}).    \label{a_od_t}
\end{equation}
Formula (\ref{a_od_t}) agrees well with the numerical solutions of Eq. (\ref{rna}). \\
In the last step we have to insert the solution (\ref{a_od_t}) into the Ansatz (\ref{ansatz}) 
considering conditions (\ref{wr}). The final expression for the field of the oscillating domain
 wall in the polynomial approximation has the form:
\begin{eqnarray}
\label{wiel}
\phi^{(0)} (t,\xi) &  = & \left[ \frac{3}{2} \left( \xi - \frac{1}{3} \xi^{3} \right) +
 \frac{3}{2} \delta \cos \Omega t \left( \xi - \xi^{3} \right) + \frac{3}{8} \delta^{2} 
\cos 2 \Omega t \left( \xi - 3 \xi^{3} \right) \right.   \nonumber \\ & & - \left. \frac{9}{8}
 \delta ^{2} \left( \xi - \frac{1}{3} \xi^{3} \right) \right] \Theta ( \xi_{0} (t) - 
\mid \xi \mid ) \nonumber \\ & & + \Theta ( \xi - \xi_{0} (t)) - \Theta ( -\xi_{0} (t) - \xi ) 
+ O( \delta^{3}),
\end{eqnarray}
where
\begin{equation}
\Omega = 2 + \frac{3}{8} \delta^{2} + O( \delta^{3}).
\label{oa}
\end{equation}
The function $\xi_{0} (t)$ can be regarded as the half-width of the domain wall in the $\xi$ 
coordinate. It is given by the formula:
\begin{equation}
\xi_{0}(t) = 1 - \delta \cos  \Omega t + \frac{3}{4} \delta^{2} +
\frac{1}{4} \delta^{2} \cos 2 \Omega t + O(\delta^{3}).  \label{ksizero}
\end{equation}
From Eq. (\ref{oa}) one can get period of oscillation of the domain wall $T = 2 \pi / \Omega$ : 
$T= \pi - O(\delta^{2})$. In the case of linear oscillations around static solution (one 
can regard this situation neglecting nonlinear terms in (\ref{rna2}) - it is admissible because 
$\Psi$ is small by definition) one gets $ \Omega = 2$ and $T = 2$. The formula (\ref{wiel}) can 
be adopted to the special, static case of domain wall solution, taking $\delta = 0$. 
It is equivalent to the $(1,0)$ central, critical point of Eq. (\ref{rna2}). The general 
solution $\phi^{(0)} (t,\xi)$ can be then treated as a small oscillation (with amplitude given 
by the parameter $\delta$) around the static solution $\phi^{(0)}_{s} (\xi)$ : 
\begin{equation}
\phi^{(0)}_{s} (\xi) = \frac{3}{2} \left( \xi - \frac{1}{3} \xi^{3} \right) \Theta (1 - \mid \xi \mid ).
\end{equation}
Accordingly to Eq. (\ref{wiel}) initially the domain wall is a bit sqeezed. Then it oscillates 
with small amplitude around the static solution.
%
%
\section{Correction to the polynomial solution}
\label{corr}
In this section we propose a simple extension of the pure polynomial solution
obtained above. Formula (\ref{wiel}) gives us the simplest, approximaty domain 
wall - type solution in our model. One can study wider range of phenomena (e.g.
nontrivial asymptotics of the domain wall solution as well as radiation, which can be emitted by 
the oscillating domain wall) considering higher - order corrections to this solution.
Our goal is
 to calculate perturbatively corrections to the zeroth order polynomial solution. Let us denote it 
by $ \phi^{(1)} (t, \xi)$. 
Thus, we can write the field satisfying Eq. (\ref{rr}) as:
\begin{equation}
\phi (t , \xi) = \phi^{(0)} (t , \xi) + \phi^{(1)} (t , \xi).  
\label{p_pola}
\end{equation}
In order to find the correction $ \phi^{(1)} (t, \xi)$ let us insert expression (\ref{p_pola}) 
into the field equation (\ref{prr}). Neglecting nonlinear terms in the field $\phi^{(1)}$ we 
get:
\begin{equation}
\partial_{\mu} \partial^{\mu} \phi^{(1)} - 2[3 \phi^{(0) 2} - 1] \phi^{(1)} = - \partial_{\mu}
 \partial^{\mu} \phi^{(0)} + 2[\phi^{(0) 2} - 1] \phi^{(0)}.     \label{nrr}
\end{equation}
We expect that global character of the domain wall solution during time evolution won't change 
because of topological stability (in the other words: the solution remains the planar domain wall
 type all the time, only small corrections can occur). That's why the linearization in 
(\ref{nrr}) may be done. Eq. (\ref{nrr}) is a linear one with source term $j(t , \xi )$ given by
 the formula:
\begin{equation}
j(t , \xi) = \frac{\partial^{2} \phi^{(0)}}{\partial t^{2}} - \frac{\partial^{2} \phi^{(0)}}
{\partial \xi^{2}} + 2 [ \phi^{(0) 2} - 1] \phi^{0}.  \label{zrodlo}
\end{equation}
As it was done for the polynomial solution $\phi^{(0)}$ we can split $j(t , \xi )$ into 
statical and time - dependent part:
\begin{equation}
j(t , \xi) = j_{s}(\xi) + j_{d}(t  ,\xi).
\end{equation}
From Eq. (\ref{wiel}) one gets:
\begin{equation}
 j_{s} (\xi) =  \left[ \frac{27}{4} \left( \xi - \frac{1}{3} \xi^{3} \right)^{3} + \xi^{3} 
\right] \Theta(1 - \mid \xi \mid) - \left( \delta ' (\xi - 1) + \delta ' (\xi + 1) \right), 
 \label{js}
\end{equation}
where prim denotes derivative with respect to $\xi $, and: \\
\\
\begin{eqnarray}
\lefteqn{j_{d} (t,\xi) = \left[ - \frac{3}{2} \delta \Omega^{2} (\xi - \xi^{3}) \cos \Omega t-3
 \delta (4 \xi - \xi^{3}) \cos \Omega t \right. } \nonumber \\ & & \left.  + \frac{81}{4} \delta
 \left( \xi - \frac{1}{3} \xi^3 \right)^{2}(\xi - \xi^{3})  \cos  \Omega t \right] 
 \Theta (\xi_{0}(t) - \mid \xi \mid) \nonumber \\ & & + \left[ \frac{27}{4} \left( \xi - 
\frac{1}{3} \xi^{3} \right) + \xi^{3} \right] \left[ \Theta ( \xi_{0}(t) -  \mid \xi \mid) - 
\Theta (1 - \mid \xi \mid) \right]  \nonumber \\  & & + \left[ \frac{3}{2} \delta \Omega^{2} 
\left( \xi - \frac{1}{3} \xi^{3} \right) \cos \Omega t - 3 \delta \, sgn \, \xi (1-3 \xi^{2}) 
\cos \Omega t \right] \delta (\xi_{0}(t) - \mid \xi \mid) \nonumber \\ & & + \frac{3}{2} sgn \, 
\xi (1-\xi^{2}) \left( \delta ( \xi_{0} (t) - \mid \xi \mid) - \delta (1 - \mid \xi \mid)\right)
 \nonumber \\ & & + \frac{3}{2} \delta (\xi - \xi^{3})  \cos \Omega t \; \delta ' (\xi_{0} (t)
 - \mid \xi \mid) \nonumber \\ & & - \frac{3}{2} \left( \xi - \frac{1}{3} \xi^{3} \right) 
\left[ \delta '  ( \xi_{0} (t) - \mid \xi \mid) - \delta ' (1 - \mid \xi \mid) \right] 
\nonumber \\ & & + \delta \Omega^{2}  \cos \Omega t \left[ \delta ( \xi + \xi_{0} (t)) - 
 \delta (\xi - \xi_{0} (t)) \right] \nonumber \\  & & - \left[ \delta ' (\xi - \xi_{0} (t)) 
- \delta ' (\xi - 1) \right] - \left[ \delta ' (\xi + \xi_{0} (t)) - \delta ' (\xi + 1) \right].
  \label{jd}
\end{eqnarray}
Eq. (\ref{nrr}) is complicated due to unpleasant form of the polynomial solution given by
Eq. (\ref{wiel}). We can simplify the operator on the left-hand side of Eq. (\ref{nrr}) noticing,
 that $\phi^{(0)}$ can be split into statical and time-dependent part, and predominant component
 $ \tanh \, \xi $ (which is exact, statical solution of the initial field equation) can be extracted
 from the static part as follows: 
\begin{equation}
\phi^{(0)} = \tanh \xi + \delta \phi^{(0)}_{s} (\xi) + \delta \phi^{(0)}_{d} (t , \xi),  
\label{podz}
\end{equation}
Inserting expression (\ref{podz}) into the evolution equation and rewriting some terms on the 
right-hand side we get finally: 
\begin{eqnarray} 
\lefteqn{\partial_{\mu} \partial^{\mu} \phi^{(1)} - 2 \left[ 3 \tanh^{2} \xi - 1 \right] 
\phi^{(1)} = j_{s}(\xi) + j_{d} (t , \xi)} \nonumber \\ & & + 6 \left( \phi^{(0)2}_{s} - 
\tanh^{2} \xi \right) \phi^{(1)} \nonumber \\ & & + 6 \, \delta \phi^{(0)}_{d} \left( \delta
 \phi^{(0)}_{d} + 2 \, \delta \phi^{(0)}_{s}+ 2 \tanh \xi \right) \phi^{(1)}.  \label{pert}
\end{eqnarray}
The component in front of $\phi^{(1)}$ in the third term on the right - hand side is nonzero 
only on the border of the domain wall and quickly tends to zero when $\mid \xi \mid \rightarrow \infty $. 
Analogously, the component in front of $ \phi^{(1)}$ in the fourth term is of order of the 
small parameter $\delta $. Thus, we have the following, strongly suggested method of solving 
Eq. (\ref{pert}): in the first step we put $ \phi^{(1)} = 0$ on the right-hand side of 
Eq. (\ref{pert}) and solve linear equation for $\phi^{(1)}$ with source term $j_{s}(\xi) + j_{d}(t , \xi)$.
 The solution obtained in this way can be inserted to the right-hand side of Eq. (\ref{pert}),
 so in the second step we have to solve equation with the same linear operator but a new source 
term. We find full solution as a result of such as iterating procedure. Nevertheless if components in
 front of $\phi^{(1)}$ on the right-hand side are small (and that is our case), we can find predominant
 part of full solution in the first step. 

Due to the form of source term (which consist of 
static and time-dependent part) it seems natural to split $\phi^{(1)}$ into two 
components: the static one and the time-dependent one. They are denoted respectively by 
$\phi^{(1)}_{s}$ and $\phi^{(1)}_{d}$. From Eq. (\ref{pert}) we get: 
\begin{equation} 
\phi^{(1) \prime \prime}_{s} - 2 \left( 3 \tanh^{2} \xi - 1 \right) \phi^{(1)}_{s}  = j_{s} (\xi) + 6 
\left( \phi^{(0)2}_{s} - \tanh^{2} \xi \right) \phi^{(1)}_{s},  \label{e1} 
\end{equation} 
for the static part of solution, and: 
\begin{eqnarray} 
\lefteqn{\Box \phi^{(1)}_{d} - 2 \left[ 3 \tanh^{2} \xi - 1 \right] \phi^{(1)}_{d} = 
j_{d}(t, \xi)} \nonumber  \\ & &+ 6 \, \delta \phi^{(0)}_{d} \left( \delta \phi^{(0)}_{d} +
2 \, \delta \phi^{(0)}_{s} + 2 \tanh \xi \right) \left( \phi^{(1)}_{d} + \phi^{(1)}_{s}
\right) \nonumber \\ & & + 6 \left( \phi^{(0)2}_{s} - \tanh^{2} \xi \right) \phi^{(1)}_{d}, 
\label{e2} 
\end{eqnarray} 
for the time-dependent part. In the last equation we use the standard notation:
\begin{equation}
\Box = - \frac{\partial^{2}}{\partial t^{2}} + \frac{\partial^{2}}{\partial \xi^{2}}.
\end{equation}
%
%
\section{Static correction to the polynomial solution}
\label{static}
Our goal in this section is to solve Eq. (\ref{e1}) and find static part of the function
$\phi^{(1)} (t , \xi) $. We adopt the method of solving, which was discussed in the previous 
section. In the first step we put $\phi^{(1)}_{s} = 0 $ on the right-hand side of Eq. 
(\ref{e1}) and solve it in a standard way as a linear, inhomogeneous equation. Then solution 
$\phi^{(1)}_s $ found in this way is inserted to the r.h.s. of Eq. (\ref{e1}) and in the 
next step equation with a new source term should be solved. Nevertheless, if expression which 
occur in a new source term as a result of this procedure is small we get pretty good approximation 
in the first step. The term in front of $\phi^{(1)}_{s}$ on the r.h.s. of Eq (\ref{e1}) is 
nonzero only on the border of the domain wall and quickly tends to zero when $\mid \xi \mid \rightarrow \infty $. 
Eventually we expect to have a good approximation in the region outside the wall. This result 
seems to be acceptable - inside the domain wall the accuracy can be improved by higher degree of 
the polynomial in Ansatz (\ref{ansatz}). Finally we are going to solve the linear equation:
\begin{equation} 
\phi^{(1) \prime \prime}_{s} - 2 \left[ 3 \tanh^{2} \xi - 1 \right] \phi^{(1)}_{s} = j_{s} (\xi).
\label{e3}
\end{equation}
The solution can be found by the standard Green's function technique. There exist two linearly 
independent solutions of the homogenous part of Eq. (\ref{e3}):
\begin{eqnarray}
f_{1} (x) & = & \frac{1}{ \cosh^{2} x}, \nonumber \\
f_{2} (x) & = & \frac{1}{8} \sinh 2x + \frac{3}{8} \tanh x +\frac{3}{8} \frac{x}{\cosh^{2} x}. 
\label{l_n}
\end{eqnarray}  
As the Green's function we take:
\begin{eqnarray}
\lefteqn{ G( \xi , x) = f_{1}(x)\,f_{2}(\xi) \Theta( \xi - x)} \nonumber \\
& & - f_{1}(\xi) \, f_{2}(x) [ \Theta( \xi - x) - \Theta( -x)]. \label{green}
\end{eqnarray}
The Green's function $G (\xi , x)$ obeys the condition:
\begin{equation}
G (\xi = 0 , x) = 0 \;\;\; \mbox{ dla }\;\;\; x \in (-\infty , \infty).
\end{equation}
The general solution of Eq. (\ref{e3}) has then the form:
\begin{equation}
\phi^{(1)}_{s} (\xi) = A\, f_{1} (\xi) + B\, f_{2} (\xi) +
\int_{R} G( \xi ,x) j_{s} (x) dx,   \label{rgreen}
\end{equation}
where $A$ and $B$ are arbitrary constants. Inserting the Green's function (\ref{green}) into 
the formula (\ref{rgreen}) we get: 
\begin{equation} 
\phi^{(1)}_{s} (\xi) = \left( A - \int_{0}^{\xi} f_{2} (x) j_{s} (x) dx \right) f_{1} (\xi) + 
\left( B + \int_{- \infty}^{\xi} f_{1} (x) j_{s} (x) dx \right) f_{2} (\xi). 
\label{rgr1}
\end{equation}
We have to put $A = 0$ and $B = 0$ because we are looking for a solution generated by source 
term, not for the homogeneous equation solution. Another reason for keeping $ B = 0$ is quick 
(exponential) growth of the function $f_{2} (x)$ for $\mid \xi \mid \rightarrow \infty $. 
In the case $B \neq 0$ the solution $\phi^{(1)}_{s} ( \xi )$ grows exponentially and doesn't 
meet requirements of the perturbative calculus. Combining expressions (\ref{js}), (\ref{green}) 
and (\ref{rgr1}) we get finally:
\begin{equation}
\phi^{(1)}_{s} (\xi > 1) = - \left[ \int_{0}^{1} f_{2} (x) \left( \frac{27}{4}
(x - \frac{1}{3} x^{3})^{3} + x^{3} \right) dx  \right] f_{1} (\xi),  \label{s1}
\end{equation}
and
\begin{eqnarray}
\lefteqn{ \phi^{(1)}_{s} (0 \leq \xi \leq 1) = \left[ \int_{0}^{\xi} f_{2} (x)
\left( \frac{27}{4} \left( x - \frac{1}{3} x^{3} \right)^{3} + x^{3} \right) dx \right]  
f_{1} (\xi)} \nonumber  \\
& &+ \left[ \int_{-1}^{\xi} f_{1} (x) \left( \frac{27}{4} \left( x - \frac{1}{3} x^{3} 
\right)^{3} + x^{3} \right)
dx  \right] f_{2} (\xi).     \label{s2}
\end{eqnarray}
Solution for $\xi < 0$ has opposite sign. \\
Integration in the formula (\ref{s1}) can be done numerically, yielding:
\begin{equation}
\phi^{(1)}_{s} (\xi > 1) = \frac{ C }{ \cosh^{2} (\xi)},
\end{equation}
where $C \simeq -0.731  $. As one can expect $\phi^{(1)}_{s} ( \xi )$ tends to zero when 
$\xi \rightarrow \infty$. The static solution  with the correction 
$ \phi^{(0)}_{s} + \phi^{(1)}_{s}$ is presented in  Fig. 2, where one can compare it with the 
pure polynomial solution $\phi^{(0)}_{s}$ and the strict, static solution $\tanh \; \xi $. 
As it is easy to notice the correction obtained above forces the polynomial solution to the 
well-known strict, static solution. It gives us a strong argument, that spliting the static 
part of polynomial solution into two parts (one of them is $ \tanh \; \xi $ ) is acceptable. 
This fact will be used in the next section. \\ 
Eq. (\ref{e1}) can be also solved in a straightforward way using numerical methods. Numerical 
analysis confirms the results obtained above.
%
%
\section{Time - dependent correction to the polynomial solution}
\label{dyn}
In this section we find time-dependent correction to the polynomial approximation.
We have to solve Eq. (\ref{e2}) using the iterative method proposed in section
\ref{corr}. The second term on the r.h.s. of the Eq. (\ref{e2}) is of the order
of the small parameter $\delta $ and can be neglected in the first approximation. 
The coefficient in front of $\phi^{(1)}_{d}$ in the third term on the r.h.s. 
is the same as in Eq. (\ref{e1}) discussed in the previous section. Now it is 
necessary to specify more precisely the scenario of evolution of the 
discussed domain wall. For $t < 0$ the domain wall is static and, as was
 discussed in the section \ref{stat} has the form of the strict, static 
solution $tanh \; \xi $ (see (\ref{stat})). In $ t = 0$ the domain wall is 
squeezed by $ 2 \delta $ as a result of action of an external force. The evolution 
at later times is governed by the field equation (\ref{e2}). \\
This scenario helps us to understand the role of the last term on the 
r.h.s. of Eq. (\ref{e2}). In the simplest case we can state that it is small
 for $\xi \rightarrow \pm \infty $, and by neglecting it we obtain good approximation
 in the first step in this region. Nevertheless from the discussion presented 
above we can conclude that we get quite good solution in the whole range of 
$ \xi $. To prove this we rewrite the third term on the l.h.s. Thus we have:
\begin{eqnarray}
\lefteqn{\Box \phi^{(1)}_{d} - 2\left[ 3 \phi^{(0) \, 2}_{s} (\xi) - 1 \right]
\phi^{(1)}_{d} = j_{d}(t, \xi)} \nonumber  \\ 
& & + 6 \, \delta \phi^{(0)}_{d} \left( \delta \phi^{(0)}_{d} + 2 \, \delta 
\phi^{(0)}_{s} + 2 \tanh \xi \right) \left( \phi^{(1)}_{d} + \phi^{(1)}_{s} 
\right).
\label{ap}
\end{eqnarray}
Due to the results of section \ref{stat} the static part of the polynomial
 solution is corrected by the function $\phi^{(1)}_{s} $, which has been calculated 
above. When we take some terms of higher order (which can be obtained in the
 next steps of calculation) we can change $\phi^{(0) \; 2}_{s}$ on the r.h.s.
 of Eq. (\ref{e2}) by "improved" expression $( \phi^{(0)}_{s} + \phi^{(1)}_{s} )^{2}$. 
From the discussion presented above and illustrated in Fig. 2 we have:
\begin{equation} \phi^{(0)}_{s} + \phi^{(1)}_{s} \approx \tanh \xi.
\end{equation}
In the first step we are then going to solve the linear, inhomogeneous equation:
\begin{equation}
\Box \phi^{(1)}_{d} - 2 \left[ 3 \tanh^{2} \xi - 1 \right] \phi^{(1)}_{d} = j_{d}(t , \xi).
\label{e4}
\end{equation}This can be rewritten as a wave equation:
\begin{equation}
\left[ \frac{\partial^{2}}{\partial t^{2}} + D^{2} \right] \phi^{(1)}_{d} (t , \xi) =
 - j_{d} (t , \xi),  \label{e5}
\end{equation}
where operator $D^{2}$ has the form:
\begin{equation}
D^{2} = - \left[ \frac{\partial^{2}}{\partial \xi^{2}} - 2(3 \tanh^{2} \xi - 1)\right].
\label{opr}
\end{equation}
Eq. (\ref{e5}) can be solved by the standard Green's function technique.
 We calculate Green's function for opearator $D^{2}$ using expansion in eigenfunctions. \\
We are looking for the Green's function $G(\xi , t ; \xi ' ,t')$, which fulfill
 the equation:
\begin{equation}
\left[ \frac{\partial^{2}}{\partial t^{2}} + D^{2} \right] G(\xi,t;\xi',t')
 = \delta(t-t') \delta(\xi - \xi'),
\end{equation}
and obeys the condition:
\begin{equation}
G(\xi,t;\xi',t') = 0 \;\;\; \mbox{  for  } \;\;\; t<t'.
\end{equation}
If the set of eigenfuctions $\{ \psi_{n} \}$ is given, we can construct Green's
function in a standard manner. Detailed description of this procedure is given
 in \cite{Barton}. The Green's function can be written as:
\begin{equation}
G(\xi,t;\xi',t') = K(\xi,t;\xi',t') \Theta (t - t'),   \label{falgr}
\end{equation}
where propagator $ K( \xi , t ; \xi' , t ) $ has the form:
\begin{equation}
K(\xi,t;\xi',t') = \psi_{0}(\xi) \psi_{0}^{*} (\xi') (t-t') +
\sum_{n \neq 0} \psi_{n} (\xi) \psi_{n}^{*} (\xi')
\frac{\sin [\sqrt{\lambda_{n}}(t-t')]}{\sqrt{\lambda_{n}}}.   
\label{falpr}
\end{equation}
The solution of the inhomogenous equation is given by the formula:
\begin{equation}
\phi^{(1)}_{d} (t,\xi) = - \int_{R} d \xi' \int_{- \infty}^{t} dt' 
K(\xi,t;\xi',t') \, j_{d} (t',\xi'). \label{fal1}
\end{equation}
Because $j_{d} (t < 0,\xi) = 0$ we can simplyfy (\ref{fal1}) as folows:
\begin{equation}
\phi^{(1)}_{d} (t,\xi) = - \int_{R} d \xi' \int_{0}^{t} dt'
K(\xi,t;\xi',t') \, j_{d} (t',\xi'). \label{fal2}
\end{equation}
The problem of calculating time-dependent correction to the polynomial solution
 is then reduced to finding the system of eigenfunctions of operator $D^{2}$. 
This system is well-known (see e.g. )and consists of two discrete eigenvalues and continous
 spectrum with corresponding eigenfunctions:
\begin{eqnarray}
\psi_{0} (x) & = & \frac{\sqrt{3}}{2} \frac{1}{\cosh^{2} x} \;\;\; \mbox{ for }
 \;\;\; \lambda_{0} = 0,       \label{wdys1} \\
\psi_{1} (x) & = & \sqrt{\frac{3}{2}} \frac{\sinh x}{\cosh^{2} x} \;\;\;
\mbox{ for } \;\;\; \lambda_{1} = 3,         \label{wdys2}
\end{eqnarray}
\\
\begin{equation}
\psi_{k} (x) = \frac{1}{\sqrt{2 \pi (k^{2}+1)(k^{2}+4)}} \, e^{ikx}
\left[ 2-\frac{3}{\cosh^{2} x} - 3ik \tanh x - k^{2} \right],
\label{wcgle}
\end{equation}
where:
\begin{displaymath}
\lambda_{k} = k^{2} + 4, \; \; \; \;  k \in R^{+}. 
\end{displaymath}
The eigenfunctions from the continous part of the spectrum correspond to the
real eigenvalues and can be split into two sets of eigenfunctions, orthogonal 
to each other:
\begin{equation}
\psi^{(1)}_{k} (x) = \frac{1}{\sqrt{2 \pi (k^{2}+1)(k^{2}+4)}}
\left[ \left( 2-k^{2} - \frac{3}{\cosh^{2} x} \right) \cos k x + 3k \sin k x 
\tanh x \right], \label{wcgle1}
\end{equation}
and:
\begin{equation}
\psi^{(2)}_{k} (x) = \frac{1}{\sqrt{2 \pi (k^{2}+1)(k^{2}+4)}}
\left[ \left( 2-k^{2} - \frac{3}{\cosh^{2} x} \right) \sin kx - 3k \cos kx 
\tanh x \right]. \label{wcgle2}
\end{equation}
\\
Inserting formulae (\ref{wdys1}) - (\ref{wcgle2}) into (\ref{falgr}) - (\ref{fal2})
we get the expression for the retarded Green's function:
\begin{eqnarray}
\lefteqn{G_{r} (\xi,t;\xi',t') = \Theta (t-t') \left( (t-t') \psi_{0}(\xi)
 \psi_{0}^{*} (\xi')  \right. } \nonumber \\ & & + \frac{1}{ \sqrt{3} } 
\sin [\sqrt{3} (t-t')] \psi_{1}(\xi) \psi_{1}^{*}(\xi') \nonumber \\ & & + 
 \left. \sum_{i = 1,2} \int_{0}^{\infty} dk \frac{\sin [ \sqrt{k^{2}+4}(t-t')]} 
{\sqrt{k^{2}+4}} \psi^{(i)}_{k} (\xi) \psi^{(i)*}_{k} (\xi') \right) .
\label{g1}
\end{eqnarray}
We get the solution $\phi^{(1)}_{d} ( t , \xi ) $ integrating
 $ G_{r} ( \xi , t ; \xi' , t') $ with the source term $ j_{d} ( t' , \xi')$
 in the whole range of $ \xi' $. As $ j_{d} ( t' , \xi')$ is odd in variable
 $ \xi'$, part of $G_{r}$ even in this variable does not give any contribution
 to the final solution. The Green's function can be rewritten as:
\begin{eqnarray}
\lefteqn{G_{r}(\xi,t;\xi',t') = \Theta (t-t') \left( \frac{1}{\sqrt{3}}
 \sin [  \sqrt{3} (t-t') ] \psi_{1} (\xi) \psi_{1}^{*} (\xi') \right. } \nonumber \\
& & + \left. \int_{0}^{\infty} dk 
\frac{\sin [ \sqrt{k^{2}+4}(t-t') ] }{\sqrt{k^{2}+4}}\psi_{k}^{(1)}(\xi) 
\psi_{k}^{(1)*} (\xi') \right) . \label{g2}
\end{eqnarray}
Finally, inserting Eq. (\ref{g2}) into (\ref{fal2}) we can write solution 
$\phi^{(1)}_{d} (t , \xi)$. For convenience we split it into two parts due to
 the parts of  Green's function (\ref{g2}). It reads:
\begin{equation}
\phi^{(1)}_{d} (t,\xi) = \phi^{(1)}_{d \, (\sim)} (t,\xi) +
 \phi^{(1)}_{d \,(-)}(t,\xi),
\end{equation}
where $\phi^{(1)}_{d \, (\sim)}$ and $\phi^{(1)}_{d \, (-)}$ are given by
formulae:
\begin{eqnarray}
\lefteqn{\phi^{(1)}_{d \, (\sim)} (t,\xi) = - \frac{1}{2 \pi} \int_{R} d 
\xi' \int_{0}^{t} dt' \int_{0}^{\infty} dk \frac{\sin [ \sqrt{k^{2}+4} (t-t') ]
 }{(k^{2}+1) (k^{2}+4)^{3/2}} \, j_{d}(t',\xi')} \nonumber \\ & &
 \left( (2-k^2) \sin k \xi - 3k \cos k \xi \tanh \xi -
 \frac{3 \cos k \xi}{\cosh^{2} \xi} \right) \nonumber \\ & &
 \left( (2-k^2) \sin k \xi' - 3k \cos k \xi' \tanh \xi' - 
 \frac{3 \cos k \xi'}{\cosh^{2}\xi'} \right),  \label{fid1}
\end{eqnarray}
\begin{equation}
\phi^{(1)}_{d \, (-)} (t, \xi) = - \frac{\sqrt{3}}{2} 
\frac{\sinh \xi}{ \cosh^{2} \xi} \int_{R} d \xi' \int_{0}^{t} dt' j_{d}
 (t',\xi' ) \sin [ \sqrt{3} (t-t') ]\frac{\sinh \xi' }{\cosh^{2} \xi' }.  
\label{fid2}
\end{equation}
\\
Let us give an interpretation of these two parts of the final solution. 
$\phi^{(1)}_{d \, (-)}$ 
is connected with the $\psi_{1}$ mode in the system of eigenfunctions of the
 operator $D^{2}$. The interaction with $\psi_{1}$ generates in the
 time-dependent correction $\phi^{(1)}_{d}$ the component, which exponentialy
 vanishes for large $\xi $ and oscillates. It can be treated as a form of 
excitation of the domain wall given by $\psi_{1} $. The crucial role which 
this component plays in dynamics of our system is more transparently visible
 in the process of collision of kinks (see \cite{zd1} and \cite{zd2}).
 \\ The component $\phi^{(1)}_{d \, (\sim)}$ has the form of a wavepacket
 - the oscillations of the domain wall generate radiation. Thus we have the 
following scenario of the discused phenomena: in $t = 0$ the external force sqeezed 
the domain wall by $ 2 \delta$. The wall oscillates around the static 
configuration. The oscillations generate radiation as in the formula 
(\ref{fid1}). It is natural to restrict $\phi^{(1)}_{d \, (\sim)} $ 
to waves going out from the wall: 
\begin{eqnarray}
\lefteqn{\phi^{(1)}_{d \, (\sim)} \rightarrow \phi^{(1)}_{d \, (\rightarrow)}
(t, \xi >0) = -\frac{1}{4 \pi} \int_{R} d\xi' \int_{0}^{t} dt'
\frac{1}{(k^{2}+1)(k^{2}+4)^{3/2}}j_{d} (t', \xi')} \nonumber \\ & & 
\cos [k( \xi - \xi') - \sqrt{k^{2}+4}(t-t')] \left[ 3k \tanh \xi 
\left( 2-k^{2} - \frac{3}{\cosh^{2} \xi'} \right) \right. \nonumber \\ & & 
\left. - 3k \tanh \xi' \left( 2-k^{2} -\frac{3}{\cosh^{2} \xi} \right) 
 \right] \nonumber \\ & & + \sin [k( \xi - \xi') - \sqrt{k^{2}+4}(t-t')] 
\left[ -2 \left( 2 - k^{2} - \frac{3}{\cosh^{2} \xi} \right) \right. 
\nonumber \\ & & 
\left. \left( 2 - k^{2} - \frac{3}{\cosh^{2} \xi'} \right) - 9 k^{2} \tanh
 \xi \tanh \xi' \right]. \label{comp} 
\end{eqnarray} 
For $ \xi < 0$ the result is analogous. \\
The solution $\phi^{(1)}_{d \, (\rightarrow)}$  has the form of a wavepacket:
\begin{equation}
\phi^{(1)}_{d \, (\rightarrow)} (t , \xi) = \int_{0}^{\infty} \; 
dk F(k,t,\xi) e^{i(k \xi - \sqrt{k^{2}+4}t)}.
\end{equation}
The waves have the frequency $\omega (k) = \sqrt{k^2 + 4}$ so they satisfy 
the dispersion relation:
\begin{equation} \omega^{2} (k) - k^{2} = 4. 
\end{equation}
It agrees well with our expectations, bacause for large $ \mid \xi \mid $ 
Eq. (\ref{e4}) reads:
\begin{equation}
\Box \phi^{(1)}_{d} - 4 \phi^{(1)}_{d} = 0.
\end{equation}
To plot the function 
$\phi^{(1)}_{d \, (\rightarrow)}$ given by the formula (\ref{comp}) 
we calculate it numerically for some fixed values of $ t $ and 
$ \xi \in (0 , 20)$, with the step $ \Delta \xi = 0.1$. The final result is
 presented in Fig. 3 - 7. One can easily see the causal character of obtained 
solution. In the region outside the 
light-cone the field $\phi^{(1)}_{d \, (\rightarrow)}$ is nearly equal to $0$. 
The fluctuations are caused mainly by computation errors. This fact was checked
 by decreasing the step of integration - the fluctuations decrease then too. 
The field $\phi^{(1)}_{d \, (\rightarrow)}$ is small - on the figures it is 
multiplied by factor $ 10^{2} $. This once again confirms the approximation used
 in our approach.
%
%
%
\section{Backreaction of the radiation}
\label{back}
In the previous section we have shown that the considered, oscillating domain wall radiates. Thus, we have 
the energy flowing out from the domain wall. We can then expect dumping of oscillations of the 
domain wall. In this section we construct a simple 
model of this process. We start from equation of energy balans:
\begin{equation}
\frac{dE^{(0)}}{dt} = - F(t),                  \label{bilans}
\end{equation}
where $E^{(0)}$ is the energy per unit area of the domain wall for the polynomial solution and $ F(t) $ is the energy 
emitted from unit square in the moment $t$. From Eq. (\ref{energia}) one can calculate $E^{(0)}$.
Neglecting terms of the second and higher order in $\delta$ we get:
\begin{equation}
E^{(0)} = E^{(0)}_{s} + E^{(0)}_{d} \delta \cos \Omega t,
\label{e0}
\end{equation}
where $E^{(0)}_{s} \approx 1.58$, $E^{(0)}_{d} \approx 0.81$. $E^{(0)}_{s}$ is the energy of the static,
polynomial domain wall solution. One can easily compare it with the energy of the strict, static
solution $\tanh \;\xi$:  $E[\tanh \; \xi ] \approx 1.33$. as we can expect the energy $E^{(0)}_{s}$ is
a bit greater than $E[ \tanh \; \xi ] $. \\
In Eq. (\ref{e0}) $E^{(0)}$ is time-dependent. It is caused by limitted accuracy
of the polynomial solution - in the case of a strict solution energy should be independent on time.
In $ t = 0 $ we have $E^{(0)}_{s} + \delta E^{(0)}_{d}$ as the energy of our configuration. For
$t>0$ it oscillates as a result of approximation in the polynomial solution, nevertheless for the 
strict solution it should be constant. Accordingly we take:
\begin{equation}
E^{(0)} = E^{(0)}(0) = E^{(0)}_{s} + \delta E^{(0)}_{d}.
\label{e01}
\end{equation}
The parameter $\delta$ describes the amplitude of the domain wall oscillations. In reality it is 
time-dependent: $\delta \rightarrow \delta (t) $ (and we expect that it 
decreases as a result of radiation emitted from the wall). The energy flux can be easily calculated from 
the energy-momentum tensor for the action (\ref{lagr}):
\begin{equation}
T^{\mu \nu} = \partial^{\mu} \phi \, \partial^{\nu} \phi + \eta^{\mu \nu} {\cal L}.
\label{te-p}
\end{equation}
The energy flux density $T^{i0}$ is given by the formula:
\begin{equation}
T^{i0} = \partial^{i} \phi \, \partial^{0} \phi.
\end{equation}
From the continuity equation we can get in a standard manner expression for $F(t)$:
\begin{equation}
F(t) = -2 \left( \partial^{0} \phi^{(1)}_{d \, (\rightarrow)} \,  \partial^{\xi}
\phi^{(1)}_{d \, (\rightarrow)} \right) \mid_{\xi = 1 + \delta}. 
\end{equation}
As $\phi^{(1)}_{d \, (\rightarrow)} \sim \delta$ we can re-define the field as follows:
\begin{equation}
\phi^{(1)}_{d \, (\rightarrow)} = \delta \tilde{\phi}^{(1)}_{d \, (\rightarrow)}.   
\end{equation}
Thus, we have:
\begin{equation}
F(t) = \delta^{2} \left( \partial^{0} \tilde{\phi}^{(1)}_{d \, ( \rightarrow) }
\, \partial^{\xi} \tilde{\phi}^{(1)}_{d \, (\rightarrow)} \right) \mid_{\xi = 1 + \delta}.
\label{fl}
\end{equation}
Collecting the formulae (\ref{bilans}), (\ref{e01}) and (\ref{fl}) one gets the following differential
equation for $\delta ( t )$:
\begin{equation}
\dot{\delta}(t) = -\delta^{2} \frac{1}{E^{(0)}_{d}}
\left( \partial^{0} \tilde{\phi}^{(1)}_{d \, (\rightarrow)}
\, \partial^{\xi} \tilde{\phi}^{(1)}_{d \, (\rightarrow)} \right) \mid_{\xi = 1 + \delta}.
\label{e02}
\end{equation}
The solution of Eq. (\ref{e02}) with the initial condition $\delta (0) = \delta$ reads:
\begin{equation}
\delta (t) = \frac{\delta (0)}{1 - \frac{\delta (0)}{E^{(0)}_{d}} \int_{0}^{t}
\left( \partial^{0} \tilde{\phi}^{(1)}_{d \, (\rightarrow)}
\, \partial^{\xi} \tilde{\phi}^{(1)}_{d \, (\rightarrow)} \right) \mid_{\xi = 1 + \delta}
dt' }.   \label{td}
\end{equation}
Eventually to find the solution $\delta (t)$ one needs the value of the field $\phi^{(1)}_{d \, ( \rightarrow) }$
on the border of the domain wall, which was calculated numerically in the previous section. \\
We have performed the numerical computations for three different values of the initial oscillation 
amplitude, namely: $ \delta (0) = 0.01, \, 0.1, \, 0.3$. The final result is presented in Figs.
8, 9 and 10. The dumping of oscillations, as we can expect, is the quickest 
when we have the greatest amplitude. \\
The plots of function $\delta (t)$ have characteristic bends at multiplicities of half-period of
oscillations $T = 2 \pi / \Omega = \pi - O (\delta^{2})$. We can interprete them as follows:
the energy is most strongly radiated from the wall when the velocity of oscillations is the greatest 
 - near the returning points the velocity is small and the dumping is slower. This coincides with the 
half-period of oscillations. \\
For small, initial amplitudes of the oscillations the dumping is very weak. In the case of
$ \delta (0) = 0.01$ during the time of approximately 4 oscillation periods the amplitude decreases
only by $ 4 \% $. For $ \delta (0) = 0.1$ and $0.3$ the corresponding values are $25 \% $ and $ 75 \% $.
These two cases can however be outside the region of application of our approximation (it demands
$ \delta (0) \ll 1 $ ).
%
%
\section{Remarks}
\label{sum}
Let us briefly summarize the main points of our work. We have derived the planar domain wall solution
with the time-dependent thickness in the polynomial approximation. The planar domain wall oscillates
with small amplitude. We have then proposed an approach enabling us to calculate the
correction to the pure polynomial solution. We have split this correction into two parts. The 
time-dependent part contains the component which is interpreted as the radiation from the 
oscillating domain wall.

There are several possible extensions and generalizations of our approach. It can be applied without
much trouble to the other field-theoretical models. In our paper we discuss only the case of
small oscillations around the static solution. As one can easily see in Fig. 1 there exist also
solutions oscillating with amplitude which is not small. This case should be discussed separately.
In accordance with our discussion in section \ref{back} we can expect very quick dumping of the
oscillations. On the other hand, we should remember that our method is based on the
linear approximation and is reliable for small amplitudes of oscillations only. 
Thus, the process of creating the kink-antikink pair is also possible.

\Large
Acknowledgments
\\
\normalsize
I would like to thank prof. H.Arod\'{z} for his help and many stimulating discussions. I am also
grateful to dr L. Hadasz for reading the paper and helpful remarks.
%
%
%
%
%

%
\newpage
\Large{Figure captions} 
\\ 
\\
\normalsize
Fig. 1. The configuration space of the equation (\ref{e1}). \\
\\
Fig. 2. The statical polynomial solution, the improved polynomial solution
and the strict solution $\tanh \xi$. \\
\\
Fig. 3. The radiation emitted by the oscillating domain wall for $t=3$. \\
\\
Fig. 4. The radiation emitted by the oscillating domain wall for $t=6$. \\
\\
Fig. 5. The radiation emitted by the oscillating domain wall for $t=9$. \\
\\
Fig. 6. The radiation emitted by the oscillating domain wall for $t=12$. \\
\\
Fig. 7. The radiation emitted by the oscillating domain wall for $t=15$. \\
\\
Fig. 8. The dumping of oscillations for the initial amplitude $ \delta = 0.01$. \\
\\
Fig. 9. The dumping of oscillations for the initial amplitude $ \delta = 0.1$. \\
\\
Fig. 10. The dumping of oscillations for the initial amplitude $ \delta =  0.3$. \\
\\
\pagestyle{empty}
\centerline{\epsfbox{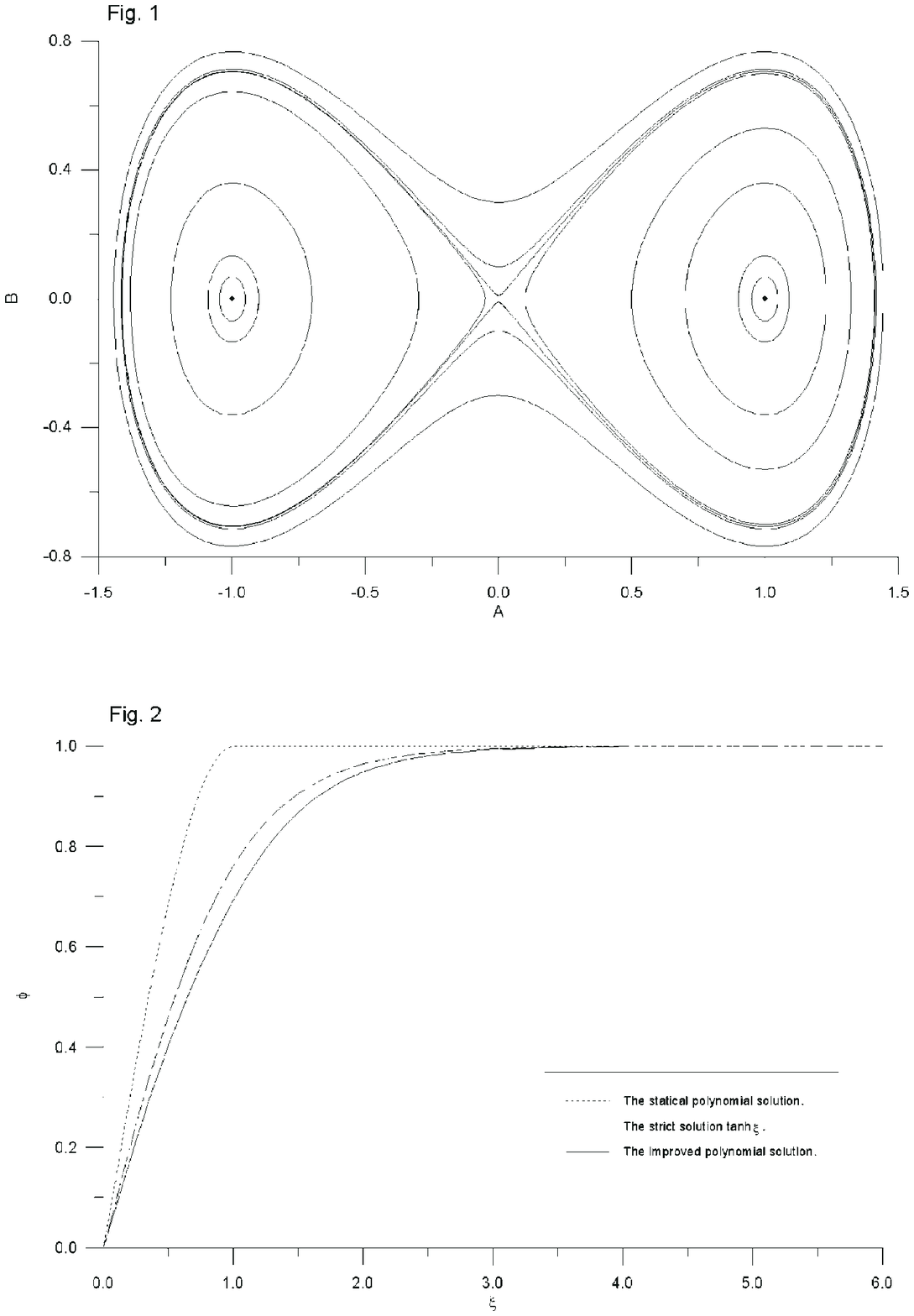}}
\centerline{\epsfbox{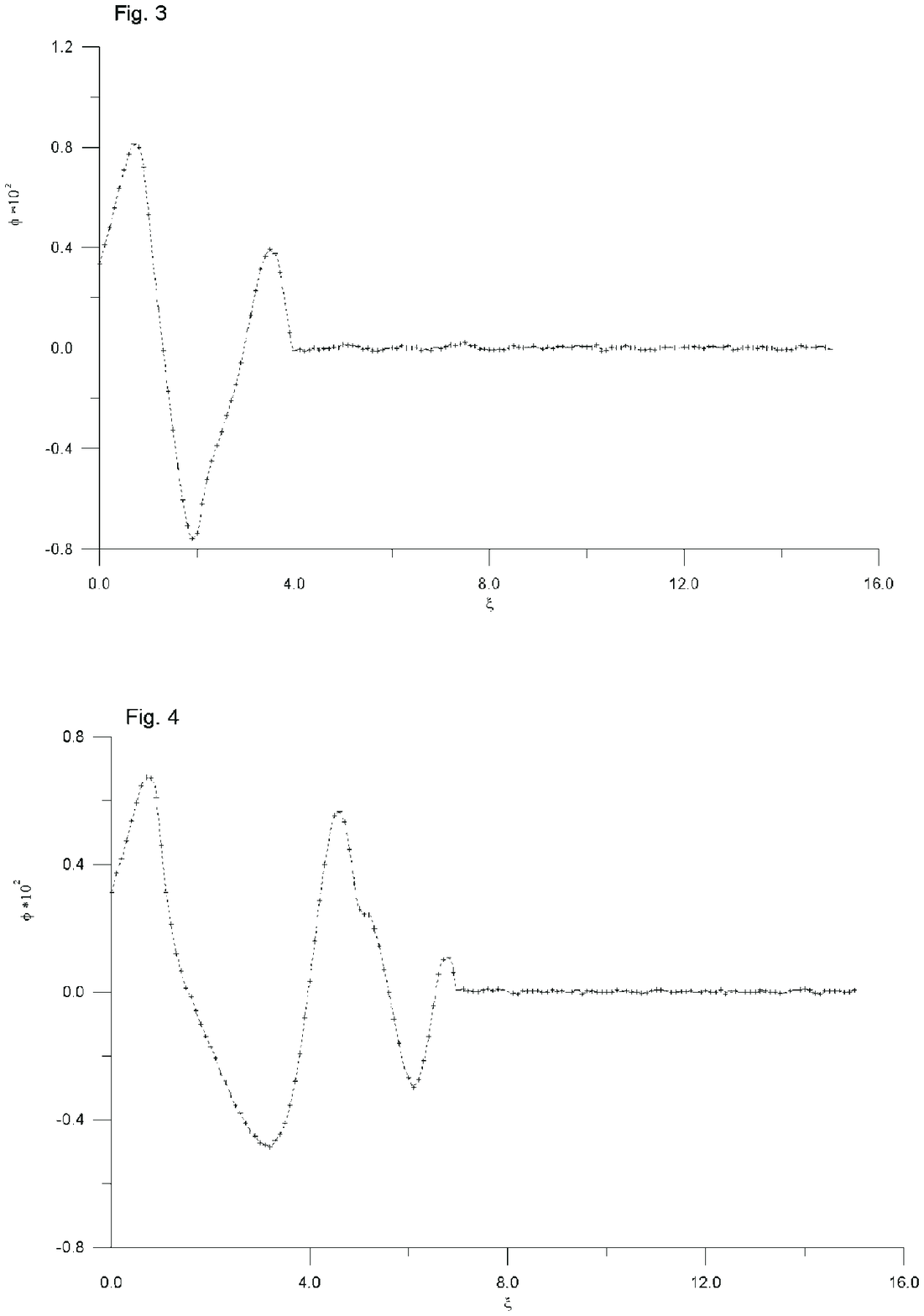}}
\centerline{\epsfbox{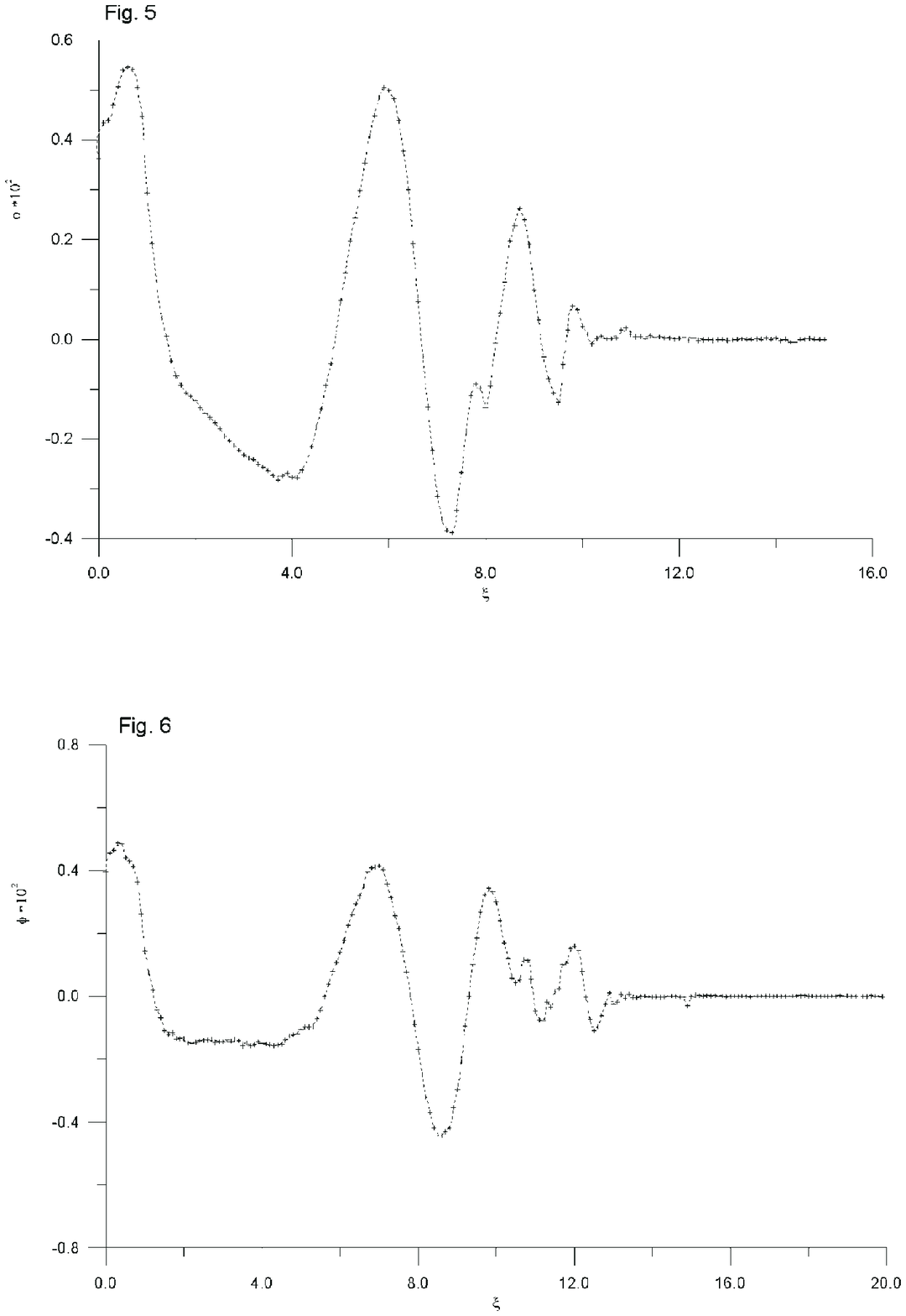}}
\centerline{\epsfbox{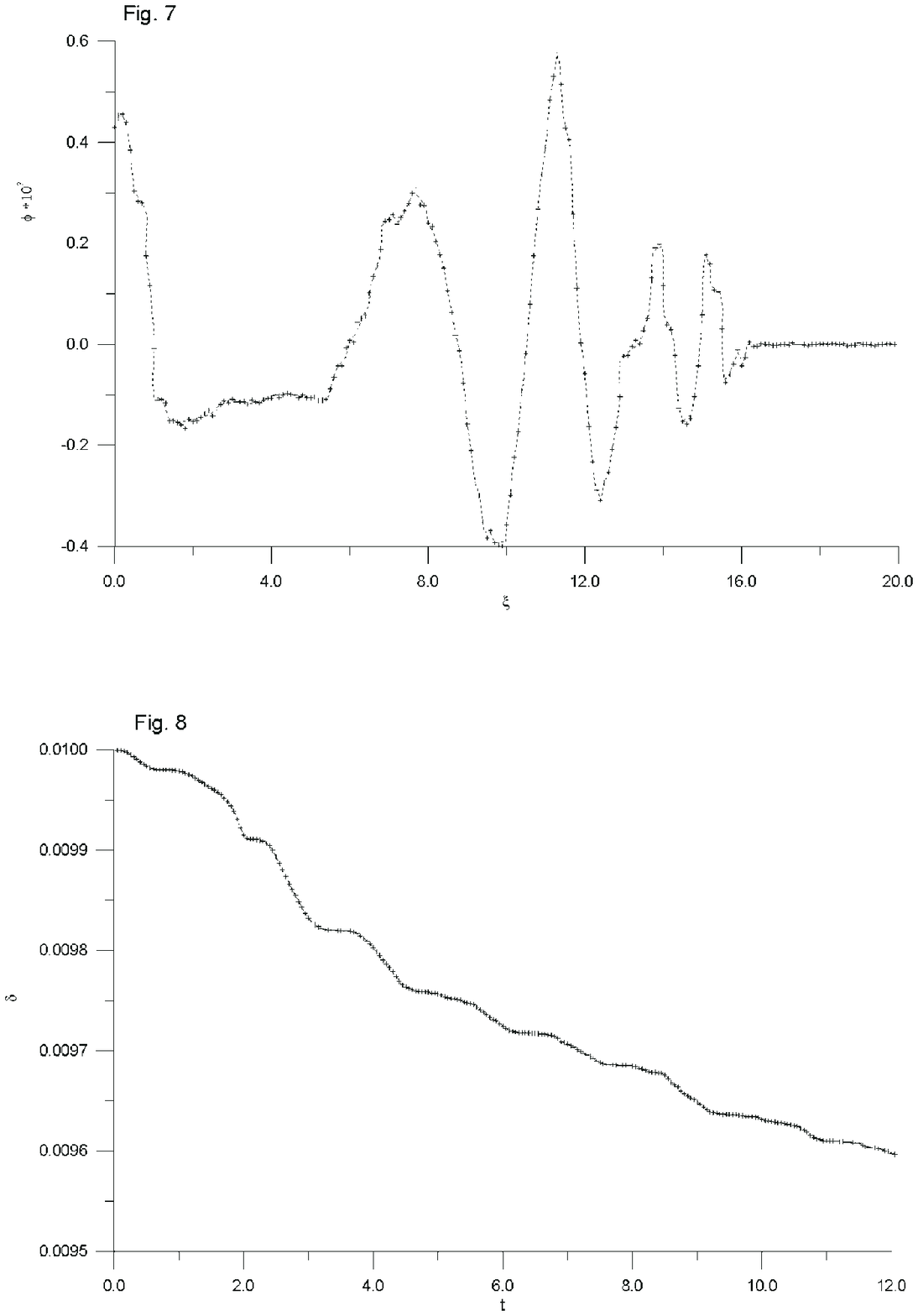}}
\centerline{\epsfbox{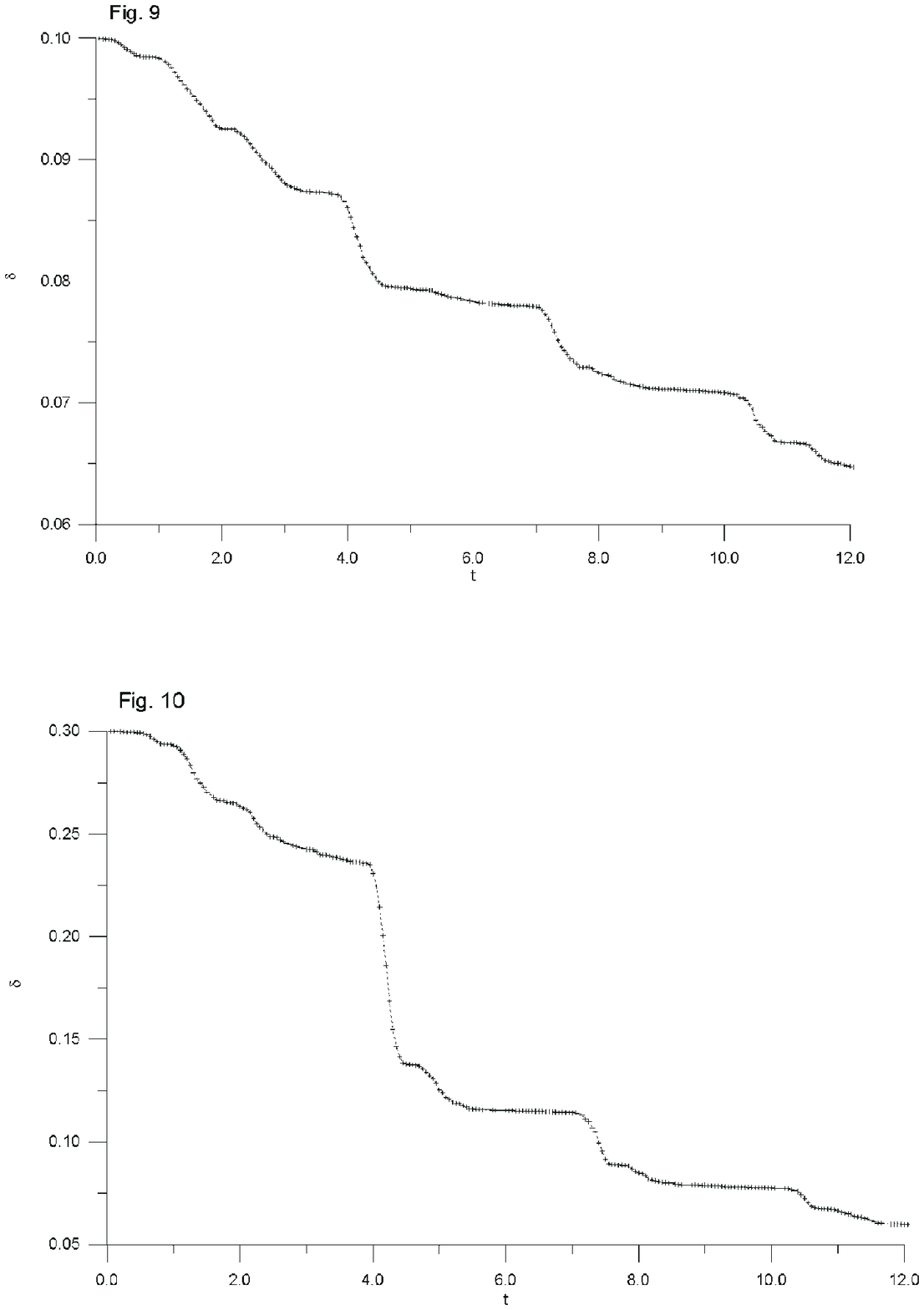}}
\end{document}